\documentclass[sigconf,table,xcdraw]{acmart}

\usepackage{bbm}
\usepackage{multirow,makecell}

\renewcommand\footnotetextcopyrightpermission[1]{}
\usepackage{fancyhdr}
\setcopyright{none}
\makeatletter
\renewcommand\@formatdoi[1]{\ignorespaces}

\makeatother
\begin{document}

\title{Audio-Visual Segmentation by \\
Exploring Cross-Modal Mutual Semantics}

\author{
Chen Liu\textsuperscript{1,2,5},
Peike Li\textsuperscript{3}, 
Xingqun Qi\textsuperscript{1,4},
Hu Zhang\textsuperscript{2}, 
Lincheng Li\textsuperscript{6}, 
Dadong Wang\textsuperscript{5}, 
Xin Yu\textsuperscript{2}
}
\affiliation{
\institution{
    \textsuperscript{1}University of Technology Sydney
    \textsuperscript{2}The University of Queensland
    \textsuperscript{3}Futureverse \\
    \textsuperscript{4}The Hong Kong University of Science and Technology
    \textsuperscript{5}CSIRO DATA61
    \textsuperscript{6}Netease Fuxi AI Lab
    }
\country{}
}
\email{{yenanliu36, xingqunqi}@gmail.com, {xin.yu, hu.zhang}@uq.edu.au}
\email{peike.li@yahoo.com, lilincheng@corp.netease.com, 
Dadong.Wang@data61.csiro.au}

\renewcommand{\shortauthors}{Trovato and Tobin, et al.}
\begin{abstract}
The audio-visual segmentation (AVS) task aims to segment sounding objects from a given video. Existing works mainly focus on fusing audio and visual features of a given video to achieve sounding object masks. 
However, we observed that prior arts are prone to segment a certain salient object in a video regardless of the audio information. 
This is because sounding objects are often the most salient ones in the AVS dataset. 
Thus, current AVS methods might fail to localize genuine sounding objects due to the dataset bias.
In this work, we present an audio-visual instance-aware segmentation approach to overcome the dataset bias.
In a nutshell, our method first localizes potential sounding objects in a video by an object segmentation network, and then associates the sounding object candidates with the given audio. 
We notice that an object could be a sounding object in one video but a silent one in another video. This would bring ambiguity in training our object segmentation network as only sounding objects have corresponding segmentation masks. We thus propose a silent object-aware segmentation objective to alleviate the ambiguity.  
Moreover, since the category information of audio is unknown, especially for multiple sounding sources, we propose to explore the audio-visual semantic correlation and then associate audio with potential objects. Specifically, we attend predicted audio category scores to potential instance masks and these scores will highlight corresponding sounding instances while suppressing inaudible ones. When we enforce the attended instance masks to resemble the ground-truth mask, we are able to establish audio-visual semantics correlation. 
Experimental results on the AVS benchmarks demonstrate that our method can effectively segment sounding objects without being biased to salient objects and also achieves state-of-the-art performance in both the single-source and multi-source scenarios.
\end{abstract}

    


\begin{teaserfigure}
\centering
\includegraphics[width=1.0\textwidth]{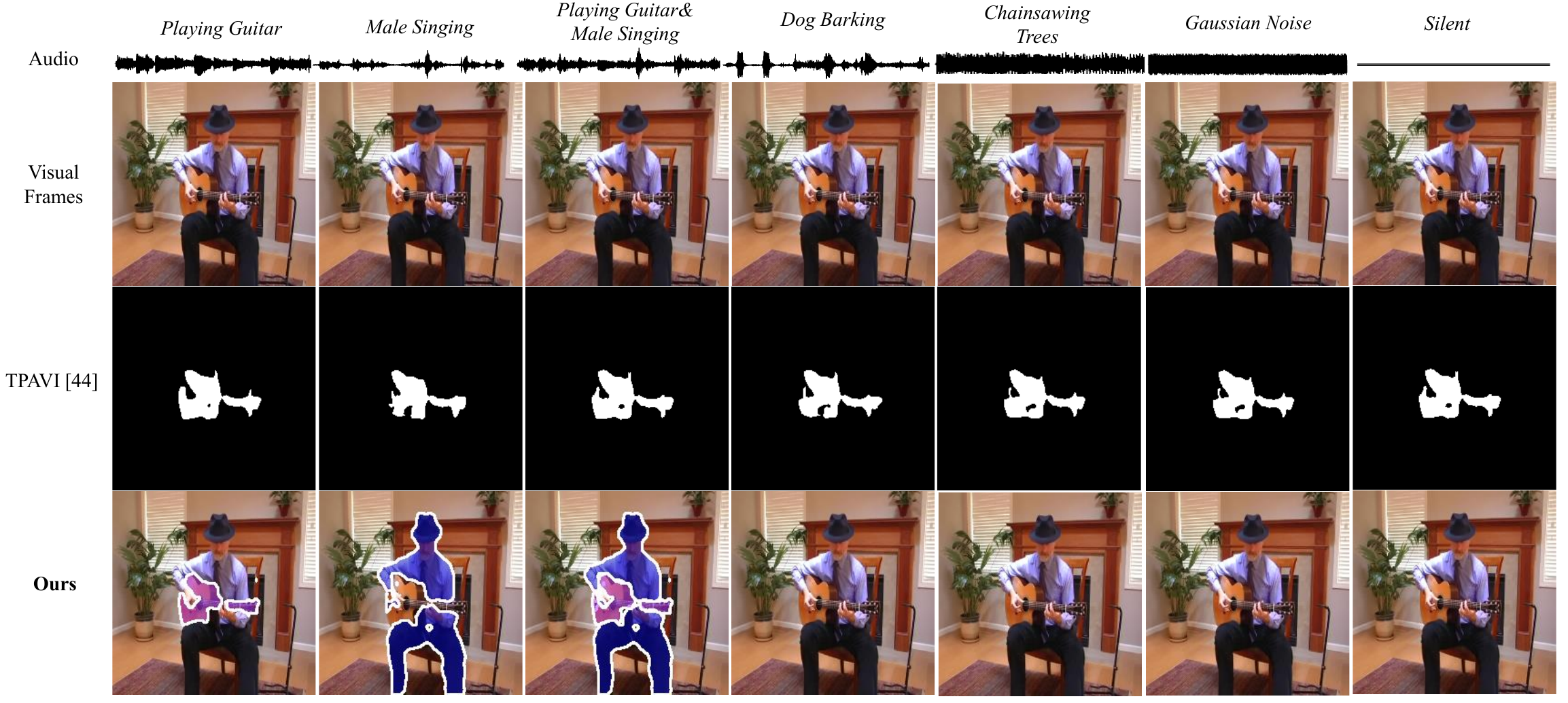}
\vspace{-4mm}
\caption{
Visual results of our method and TPAVI \cite{zhou2022audio}. Our method accurately segments sounding objects according to different audio signals. For the audio signal mixed by "Male Singing" and "Playing Guitar", our method even achieves instance-level sounding object segmentation. Here, we employ different colored masks to represent segmentation masks of different instances. Besides, our method is not misled when the guitar signal is muted or replaced by "Dog Barking", "Chainsawing Trees", or "Gaussian Noise". In contrast, the audio semantic information fails to modulate segmentation results in \cite{zhou2022audio}. No matter what the audio signal is, TPAVI always segments the most prominent guitar.}.
\label{fig:motivation}
\end{teaserfigure}

\maketitle

\section{INTRODUCTION}
Sounding source localization has been applied in various multi-media applications, such as robotic navigation~\cite{younes2023catch}, security monitoring~\cite{zhang2022recent}, wildlife conservation~\cite{schneider2021localize}, and industrial maintenance~\cite{henze2019audioforesight}. 
The Audio-Visual Segmentation (AVS) task \cite{zhou2022audio} further pushes the frontier of sounding source localization and strives to localize the masks of sounding objects in a pixel-wise manner.

Existing audio-visual segmentation methods predominantly first fuse audio-visual features and then predict sounding object masks. However, we found that current methods are prone to segment a certain salient object in a video regardless of audio information. 
For instance, as illustrated in Figure~\ref{fig:motivation}, when we replace the original audio (\emph{i.e.}, playing guitar) with the audio of another object (\emph{i.e.}, male singing), the segmentation result of \cite{zhou2022audio} remains almost the same. 
Furthermore, when the audio is muted or replaced by ``dog barking'', a blank mask is expected but \cite{zhou2022audio} still outputs the mask of the guitar. 
This implies that audio information might not play its role effectively in existing AVS approaches.
As a sounding object is often the most salient object in a video, this bias leads current AVS networks to localize a specific salient object while neglecting other potential audible objects. 
Therefore, current AVS methods would fail to segment sounding objects in accordance with different audio information.

To tackle the aforementioned issues, we introduce an audio-visual instance-aware segmentation approach. 
Unlike previous works that focus on fusing audio and visual features and then predicting segmentation masks, our method firstly attempts to localize all the potential sounding objects by learning an instance segmentation neural network. 
However, we notice that an object could be a sounding object in one video but is an inaudible one in another video but the AVS dataset only provides the segmentation masks of sounding objects. 
This would lead to ambiguity in learning the segmentation network, and the segmentation network may fail to localize all potential-sounding object masks.
To mitigate this issue, we propose a silent object-aware segmentation loss. 
It not only supervises sounding object segmentation but also can tolerate other object masks which might be potential sounding objects in the dataset. 
In this way, we can segment all potential sounding objects from a video. 

After obtaining sounding object candidates, we seek to associate the audio with genuine sounding instances. 
Note that the category information of audio is unknown, especially for multiple-sounding sources. 
Thus, we develop an audio-visual semantic correlation (AVSC) module to mine the audio and visual correspondences. 
To be specific, our AVSC first estimates category probabilities of audio via some multi-layer perceptron (MLP) layers, and then attends estimated probabilities to their corresponding candidate masks to produce a sounding object probability map. 
In learning AVSC, we force the sounding object probability map to be similar to the ground-truth mask. 
If an object instance is a sounding one, its audio category probability will be encouraged to be high. 
On the other hand, if an object is inaudible, its estimated audio category probability will be restrained. 
In this fashion, we effectively associate audio information with sounding objects while significantly alleviating over-fitting to certain salient objects.

Extensive experiments on the widely-used AVS benchmark \cite{zhou2022audio} demonstrate that our method achieves state-of-the-art performance on both single-source and multi-source audio-visual segmentation tasks. 
More importantly, we also illustrate that our segmentation results can be controlled by various audio signals, demonstrating our method is aware of audio information in segmenting sounding objects. 
In summary, our contributions are threefold:
\begin{itemize}
\item We first present an instance-aware audio-visual segmentation method that allows the segmentation results to be controlled by different audio signals, rather than overfitting to certain salient objects in the dataset. 
\item We introduce a silent object-aware segmentation objective to alleviate the segmentation supervision provided by the existing AVS dataset. With this objective, we can effectively segment potential sounding instances.
\item We design an audio-visual semantic correlation (AVSC) module to establish the association between audio and visual information. Our proposed AVSC enables us to identify the sounding instances from all the candidates without knowing the category information of audio sources.
\end{itemize}

\section{RELATED WORKS}
\noindent\textbf{Visual Sound Localization}
Sounding source localization methods aim to locate the regions in the visual frames related to the corresponding audio signals.
When the pixel-level annotations of sounding object locations are not available, previous methods mainly employ coarse heat maps or bounding boxes to localize sounding objects 
~\cite{senocak2018learning, zhao2018sound, shi2022unsupervised,zhou2023seco, zhou2023exploiting, mo2022closer, mo2022localizing, senocak2022less, chen2021localizing, song2022self, liu2022visual, oya2020we, qian2020multiple, senocak2022learning, hu2020discriminative, zhou2022contrastive}.
Those methods mainly first establish correspondences between audio and visual representations through contrastive learning ~\cite{chen2021localizing, mo2022closer, mo2022localizing, song2022self, senocak2022learning} and then localize objects based on the audio-visual feature similarity. 
For example, ~\cite{chen2021localizing} employs contrastive learning with hard negative mining to learn co-occurrence between audio and images, while ~\cite{senocak2022learning} proposes to leverage hard positive samples to align visual and audio features.
The work~\cite{qian2020multiple} further develops a multi-instance contrastive learning fashion to match audio to video frames.
However, these methods usually localize visible sound sources but struggle to identify negative cases, where sounding objects are not visible. 
Apart from the difficulty of identifying invisible sounding objects, several methods ~\cite{hu2020discriminative, liu2022visual} require knowing the number of sound sources in advance. However, this requirement might be not met, thus limiting their applications in practice.

\begin{figure*}[t] 
\begin{center}  
	\includegraphics[width=1.0\linewidth]{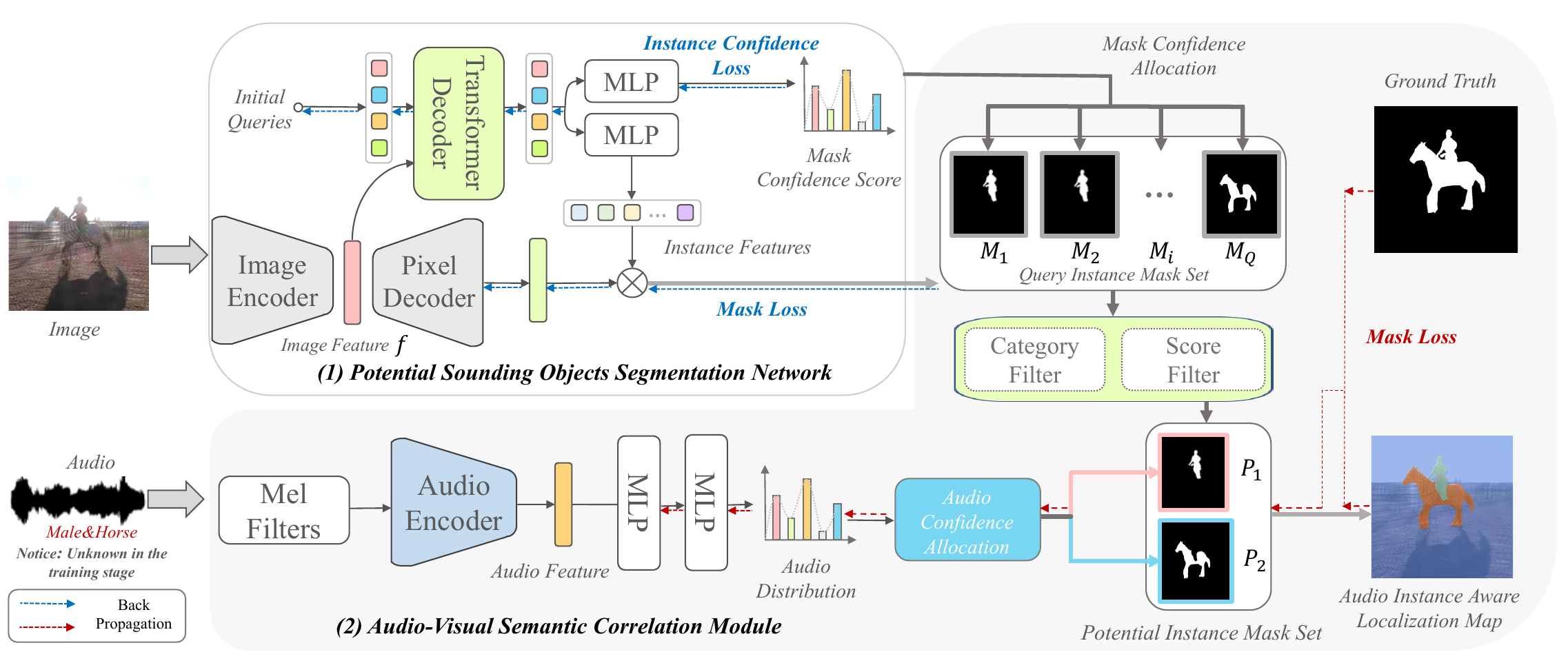} 
	\caption{Overview of our audio-visual instance-aware segmentation method. It consists of a potential sounding object segmentation network and an audio-visual semantic correlation module (AVSC). The segmentation network localizes potential-sounding objects. Then AVSC associates the audio with potential sounding instances by attending our estimated audio category labels to instance segmentation masks. During training, sounding objects' masks will be highlighted while silent ones will be suppressed.}

\label{fig2}
\end{center}  
\end{figure*}

Since previous methods cannot localize sounding objects accurately, \cite{zhou2022audio, liu2022visual} first proposes an audio-visual segmentation dataset with pixel-level audio-visual annotation, allowing researchers to localize sounding objects more precisely. 
Our work also focuses on predicting sounding object masks while aiming to achieve the ability to recognize invisible sounding sources and fully leverage audio information in segmentation.  

\vspace{0.5em}
\noindent\textbf{Guided Segmentation Networks with Conditional Information.}
Semantic or instance segmentation tasks play critical roles in computer vision. They both aim to identify objects and their boundaries within an image ~\cite{chen2018encoder, zhao2017pyramid}. 
Convolutional neural network (CNN) models have been widely used for these tasks with popular architectures, such as U-Net ~\cite{ronneberger2015u}, Mask R-CNN ~\cite{he2017mask}, and DeepLab ~\cite{chen2017deeplab,chen2017rethinking}.
Recently, transformer-based models have also shown promising results for segmentation tasks. Transformers ~\cite{vaswani2017attention}, originally proposed for natural language processing, have shown effectiveness in capturing long-range dependencies and contextual information in images ~\cite{qi2023diverse, qi2023emotiongesture}. 
Some examples of transformer-based segmentation models include DETR ~\cite{carion2020end}, SETR ~\cite{strudel2021segmenter}, and MaskFormer ~\cite{cheng2021per}. 
In addition to these model architectures, researchers have also investigated using different contextual information, as additional conditions, to improve segmentation performance. 
For instance, text ~\cite{luddecke2022image} can be used to specify the desired output segmentation masks, depth ~\cite{wang2016learning} or surface normal maps ~\cite{fan2020sne} can be used as input features, and audio ~\cite{zhao2018sound, senocak2018learning} provides additional contextual information to localize specific sounding instances. 

Audio-guided segmentation aims to leverage audio signals to guide the segmentation process.
Previous works design various mechanisms ~\cite{zhou2022audio, shi2022unsupervised, shuo2021vision, liu2022exploiting} to combine information from audio and visual information and then perform segmentation from the fused audio-visual features.
Although the audio-guided segmentation method has shown promising results ~\cite{zhou2022audio}, it cannot distinguish whether the sound comes from one object or multiple ones.
Thus, they cannot further divide a semantic segmentation mask into instance-level masks.
Furthermore, due to the bias of the current dataset \cite{zhou2022audio}, audio-guided segmentation might favor segmenting salient objects while ignoring audio information.
Thus, audio cannot play its role during segmentation. 
In contrast, the segmentation results of our proposed method can reflect different audio sources.
Moreover, even though we do not have instance segmentation masks, we can provide instance-level segmentation when multiple sound sources are given.

\section{PROPOSED METHOD}
In the audio-visual segmentation (AVS) task, our target is to segment the corresponding sounding objects in one visual frame $v$ given an audio $a$.
To solve this challenging problem, we propose an instance-aware audio-visual segmentation framework by exploring cross-modal mutual correlations between visual frames and audio signals.
As illustrated in Figure~\ref{fig2}, our framework is composed of two stages, \emph{i.e.}, a potential sounding object segmentation module and an audio-visual semantic correlation (AVSC) module.

\begin{figure}[t]
	\begin{center}  
		\includegraphics[width=0.95\linewidth]{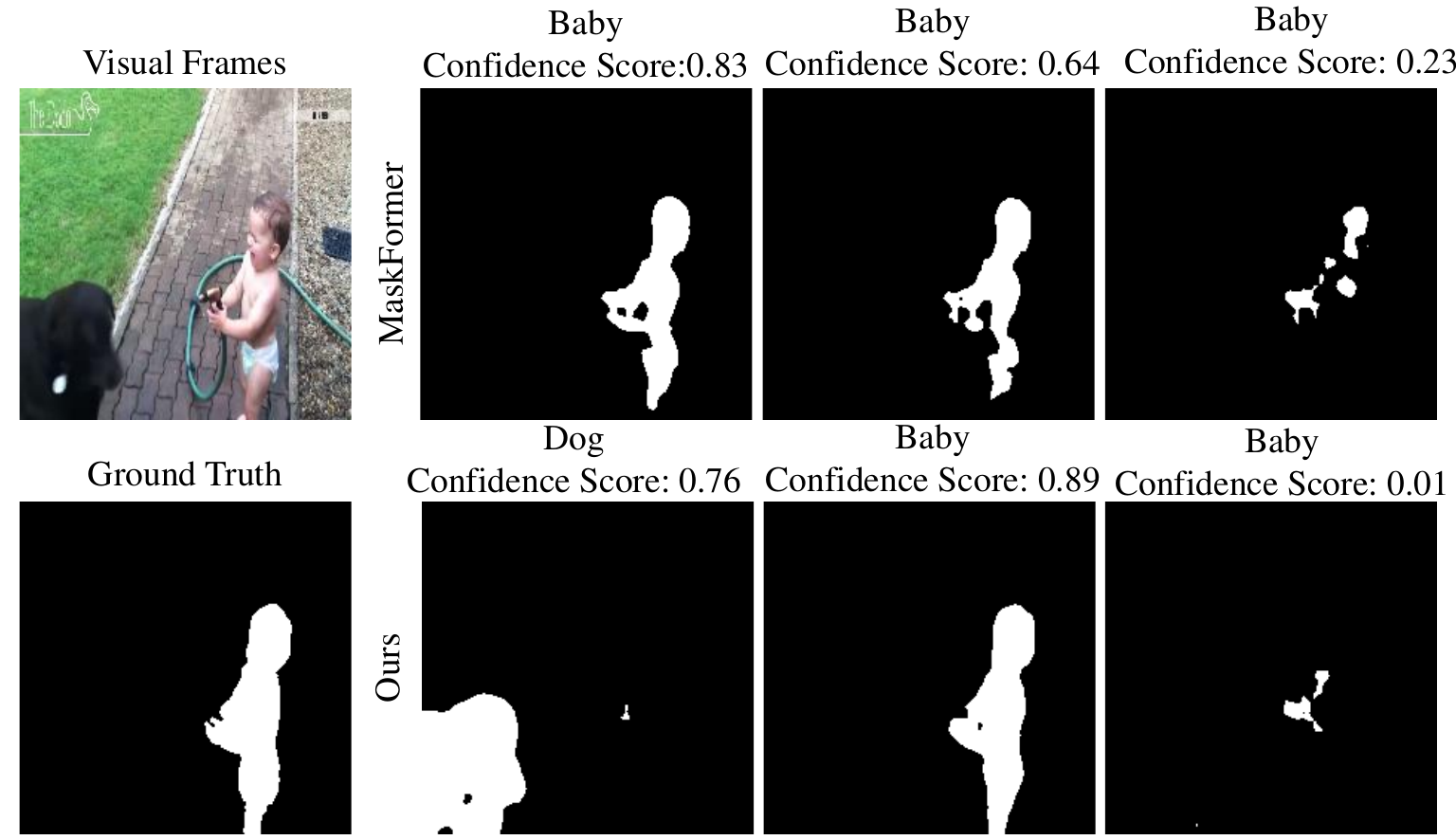} 
	\end{center}
 \vspace{-1.em}
\caption{Visualization of instance masks obtained by MaskFormer and our segmentation network.
Though we query 100 instance masks for both MaskFormer and our segmentation network, MaskFormer tends to generate masks only on one object while ours generates queries at various regions thanks to our silent object-aware segmentation loss.}
\vspace{-1.5em}
	\label{Fig_wl} 
\end{figure}

\subsection{Potential Sounding Object Segmentation}
\label{sec:3.1}

We propose the potential sounding object segmentation module to enumerate the potential sounding instances in video frames.
However, due to the differences between the typical segmentation datasets and the audio-visual segmentation dataset, we cannot directly utilize the semantic segmentation models or panoptic segmentation models to achieve this goal.
The difficulty is mainly induced by the specificity of the audio-visual segmentation dataset.
First, different from the existing typical segmentation dataset in which each pixel has a category label, \emph{e.g.}, COCO  \cite{kirillov2019panoptic, lin2014microsoft} and ADE20K \cite{zhou2017scene}, the ground-truth labels in the audio-visual dataset are just binary masks (\emph{i.e.}, 1 represents sounding and 0 represents silent) even though a video frame involves multiple sounding objects.
Second, there are nearly 89\% of the video frames that only contain one sounding object.
In other words, most of the ground truths only contain one segmentation object, while the COCO or ADE20K usually have multiple instances in an image.
This would render a segmentation network to become a saliency segmentation framework easily, failing to propose various instances in a frame.

To solve the above problems, we employ a potential sounding object segmentation network, and the architecture design inherits the spirits of the MaskFormer series~\cite{cheng2021per, cheng2022masked}.
As shown in Figure \ref{fig2}, our network is mainly composed of three components: an image encoder, a pixel decoder, and a transformer decoder.
The learnable queries in the transformer decoder represent the embedding features of instances and will be mapped to binary masks and their correspondence classification scores.
Note that even though we adopt the architecture of MaskFormer ~\cite{cheng2022masked}, directly training the network with the ground-truth will lead to inferior segmentation results, as illustrated in Figure~\ref{Fig_wl}. 
Therefore, we introduce our salient object-aware segmentation objective to train this network in Section \ref{sec:3.4}, and then obtain the set of potential sounding objects even without pixel-wise label annotations.

In detail, we feed $v \in \mathbb{R}^{3\times H\times W}$ into the segmentation network, where $H$ and $W$ refer to the height and width of the visual frame, respectively.
After the image encoder, the image feature $f$ is sent into the transformer decoder and the pixel decoder, respectively.
The pixel decoder decodes $f$ into an image pixel feature.
The transformer decoder outputs the embedding vectors of query instances, and the embedding vectors are further mapped into the final predicted results $q=\{(p_i, m_i)\}_{i=1}^{N}$,  where $p_i \in \Delta^{K + 1}$ denotes classification confidence scores of masks, and $m_i \in \{0,1\}^{H \times W}$ represents the binary masks.
Here, $K$ is the number of categories in the dataset.
Besides, we also set a non-object category to represent the background regions.
Hence the total number of categories is $K + 1$.

To mine the potential sounding objects as much as possible, the number of the predicted masks $N$ (\emph{i.e.}, queries) should be larger than that of the ground-truth $N_{gt}$.
Here, we formulate the ground-truth segments as $z_{gt} = \{(c_j, m_j)|c_j \in \{1, \dots, K\}, m_j \in \{0, 1\}^{H \times W} \}_{j=1}^{N_{gt}}$, where $m_j$ and $c_j$ represent a binary mask and the category label of a sounding object, respectively.
During training, we utilize the bipartite matching \cite{cheng2021per} to obtain the matching index $\sigma$ between the predicted results and the ground-truth, and then calculate the training loss.
After training, the potential sounding object segmentation network outputs the potential instance masks and their corresponding confidence scores.
In this fashion, we could obtain the $\emph{Query Instance Mask Set}$ which contains the masks of instances and the confidence score of each mask.
They will be employed as prior guidance to achieve the audio-visual semantic correlation in the next stage.

\subsection{Audio Feature Extraction}
\label{sec:3.2}
To accomplish instance-aware audio segmentation, we need to establish a robust semantic association between the audio and visual modalities.
Consequently, our primary focus lies in ensuring that the extracted audio features are semantically discriminative. 
Here, we employ the Bidirectional Encoder Audio Transformer (BEATs) \cite{chen2022beats}, an audio pre-training model that utilizes acoustic tokenizers to enhance audio semantics, to extract pertinent audio features.

As depicted in Figure~\ref{fig2}, we first feed the Mel-Filters transformed audio signals into the pre-trained audio encoder.
Subsequently, we apply multi-layer perceptron (MLP) blocks to obtain the audio semantic distribution $P_a \in \Delta^{K}$ over K categories. 
In this context, $\Delta^{K}$ represents the K-dimensional probability simplex.
Note that our approach does not possess knowledge of the category labels of each input audio. 
Instead, we only know the number of audio categories within the dataset.
Hence, we formulate the instance information as the category prior guidance to guarantee the semantic distribution of the audio.
To be specific, we assume that the class distribution of instances is consistent with the audio distribution.
Then, we regard the probability of the audio category corresponding to the instance category id as the probability of the object making a sound.
In this fashion, we could obtain a combined mask that indicates the sounding probability of each potential instance, and the ground-truth mask is the supervision to suppress the probability of silent instances and encourage the probability of sounding instances.
This is the distinct difference from existing sounding object localization approaches under an audio-label supervised paradigm. 
The implementation details will be explained in Section ~\ref{sec:3.3}.

\vspace{-1.0 em}

\subsection{Audio Visual Mutual Semantics Alignment}
\label{sec:3.3}
To ensure audio-visual semantic alignment, we incorporate the audio distribution into the instance masks. 
Specifically, under the assumption that the instance semantic distribution is the same as the audio semantic distribution, we could regard the audio category probability as the sounding probability of each instance and further obtain the audio instance aware localization map.
However, the queried instance mask set procured from our segmentation network might exhibit a significant level of noise. 
For example, numerous masks encompass the same instance but with varying quality, and some instance masks are even classified into the non-object category. 
Consequently, we design a category filter and a score filter to eliminate invalid segmentation.
Given the predicted results $q = {(p_i, m_i)}_{i=1}^N$, the category filter removes segmentation results belonging to the non-object category.
Subsequently, the score filter selects the instance mask with the highest mask confidence $p_i$ within each category.
Here, we would like to emphasize that the current dataset does not contain two instances of the same category producing sound simultaneously, and this case might need extra information, such as text guidance, to localize the sounding one.

Afterwards, we obtain the \emph{Potential Instance Mask Set} $q_r=\{c_i, m_i\}_{i=1}^{N_p}$, where $c_i \in \{0,1\}^{K}$ is the one-hot vector of a certain category, and  $m_i \in \{0,1\}^{H \times W}$ is the binary mask of each instance. 
Then, we multiply the instance masks with their corresponding audio distribution scores $p_a \in [0,1]$ to obtain an audio semantics-aware localization map $\mathcal{S}_{asl}\in\mathcal{R}^{H \times W}$:

\begin{equation}
    \mathcal{S}_{asl} = \sum\nolimits_{j=1}^{N_p}p_{c_j} \cdot m_j.
\end{equation}

The audio semantics-aware localization map $\mathcal{S}_{asl}$ indicates the potential locations of sounding objects.
To establish the connection between audio and visual representations, we introduce an audio-visual correspondence loss $\mathcal{L}_{avc}$.
If the audio instance aware localization map lies in or resembles the ground-truth mask, our loss will encourage the corresponding category score of the audio to be higher. Otherwise, the loss will penalize the audio category score.
In this way, we can effectively reduce the impact of silent objects from the potential sounding object set.

\subsection{Training Objective}
\label{sec:3.4}

\textbf{Training Objectives for Segmentation Network.} For the potential sounding objects segmentation network, our objective is to acquire the potential instance masks and the confidence scores associated with each instance. 
Here, we first employ an objective $\mathcal{L}_{mask\_cls}$, composed of a cross-entropy classification loss and a binary mask loss $\mathcal{L}_{mask}$. 
The binary mask loss involves a focal loss \cite{lin2017focal} and a dice loss \cite{milletari2016v}.
$\mathcal{L}_{mask\_cls}$ is defined as follows:
\vspace{-0.5em}

\begin{equation}
\mathcal{L}_{mask\_cls} (q, z_{gt}) \!= \!\sum\nolimits_{j=1}^{N} [-log p_{\sigma(j)}(c_{gt}) + \mathcal{L}_{mask}(m_{\sigma(j)}, m_{gt})],
\end{equation}

\vspace{-0.8em}
\begin{equation}
\begin{aligned}
\mathcal{L}_{mask}(m, m_{gt}) = \lambda_{f}\mathcal{L}_{Focal}(m, m_{gt}) + \lambda_{d}\mathcal{L}_{Dice}(m, m_{gt}),
\end{aligned}
\end{equation}
where the $\lambda_f$ and $\lambda_d$ are hyper-parameters and set to 20 and 1, respectively.

Moreover, to adapt the segmentation network to the audio-visual segmentation task, we design a silent object-aware segmentation loss $\mathcal{L}_{soas}$ by reducing the overlapping regions between the non-object regions and the foreground regions, expressed by:
\begin{equation}
\begin{split}
\mathcal{L}_{soas}(m_{n}, m_{gt}) = \sum\nolimits_{j=1}^{N_n}\frac{m_{n_j} \cap (\bigcup_{k=1}^{N_{gt}}m_{gt_k})}{m_{n_j} \cup (\bigcup_{k=1}^{N_{gt}}m_{gt_k})},
\end{split}
\end{equation}
where $m_{n_j}$ and $m_{gt_k}$ are the predicted mask belonging to the non-object set and ground truth set respectively, and $N_n$ and $N_{gt}$ indicates the quantity of the non-object set and ground truth set.

To sum up, the overall loss function for the potential sounding objects segmentation network is defined by:
\begin{equation}
\mathcal{L}_{sn} = \mathcal{L}_{mask\_{cls}} + \lambda_{soas}\mathcal{L}_{soas},
\end{equation}
where $\lambda_{soas}$ is empirically set to 1.

\textbf{Training Objectives for AVSC.}
We adopt a binary cross entropy loss (BCE) $\mathcal{L}_{avc}$ to supervise the audio-visual semantic correlation module, formulated as:
\begin{equation}
\mathcal{L}_{avc} = BCE(\mathcal{S}_{asl}, M_{gt}),
\end{equation}
where $M_{gt}$ indicates the ground truth mask of the audio-visual input pair.

\section{Experiments}
\subsection{Implementation Details}
\textbf{Potential sounding objects segmentation network.} We employ the swin-transformer \cite{liu2021swin} pretrained on COCO \cite{kirillov2019panoptic, lin2014microsoft} as our backbone network.
The visual input is an image of size 224 $\times$ 224 $\times$ 3 pixels.
The potential sounding objects segmentation network is trained using the Adam optimizer with a learning rate of 1e-4 and a batch size of 32. 

\vspace{0.5 em}
\noindent
\textbf{Audio-Visual Semantic Correlation Module.} We adopt BEATs \cite{chen2022beats} pretrained on AudioSet \cite{45857} as our audio extractor.
To achieve the audio distribution, we further employ some MLP layers.
The input of the audio branch is an audio signal clip with a duration of 1 second. 
The sampling rate of each raw waveform is 16,000.
After Mel-Filters, the audio signal is converted into spectrograms with 98 $\times$ 128 dimensions.
In this stage, we adopt Adam optimizer with a learning rate of 0.001 and a batch size of 64.
More details about the model configuration and data processing are provided in the supplementary material.

\begin{figure*}[t]
	\begin{center}  
		\includegraphics[width=0.95\linewidth]{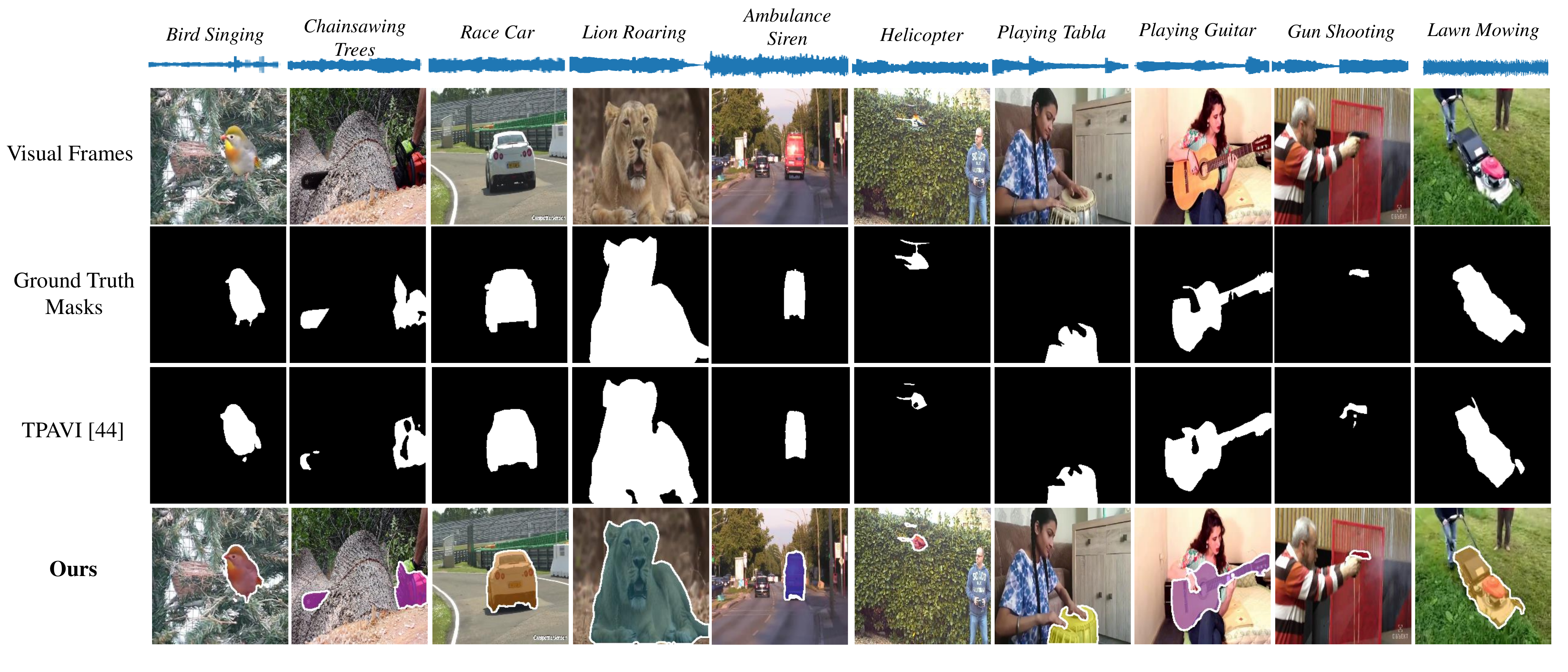}
	\end{center}
	\vspace{-1.5em}
	\caption{Qualitative comparisons with the state-of-the-art TPAVI on the single-source dataset.}
 \vspace{-0.5em}
	\label{Fig3} 
\end{figure*}

\begin{table}[t]
\centering
\caption{Quantitative comparisons with the state-of-the-art on the Single-source and Multi-source sub-datasets. Results of Jaccard index ($\mathcal{J}$) and the F-score ($\mathcal{F}$) are reported.}
\vspace{-0.5em}
\label{table1}
\small
\setlength{\tabcolsep}{4.0mm}{%
\renewcommand{\arraystretch}{1.2} 
\begin{tabular}{ccccc}
\toprule
                           & \multicolumn{2}{c}{TPAVI~\cite{zhou2022audio}} & \multicolumn{2}{c}{\textbf{Ours}}                                               \\ \cline{2-5} 
\multirow{-2}{*}{Settings} & $\mathcal{J}$ $\uparrow$           & $\mathcal{F}$ $\uparrow$           & $\mathcal{J}$ $\uparrow$                                      & $\mathcal{F}$ $\uparrow$                                      \\ \midrule
Single-source              & 78.7        & 87.9        & \cellcolor[HTML]{ECF4FF}\textbf{81.29} & \cellcolor[HTML]{ECF4FF}\textbf{88.59} \\
Multi-source               & 54.0        & 64.5        & \cellcolor[HTML]{ECF4FF}\textbf{59.5}  & \cellcolor[HTML]{ECF4FF}\textbf{65.74} \\ \bottomrule
\end{tabular}%
}
\vspace{-0.5em}
\end{table}

\subsection{Dataset and Evaluation Metrics}
\textbf{AVSBench Dataset.} 
The AVSBench dataset contains 5,356 video samples, distributed across 23 distinct auditory categories ~\cite{zhou2022audio}. 
Each video has a duration of 5 seconds and is uniformly segmented into five clips. 
Annotations for the sounding objects are provided by the binary mask of the final frame of each video clip.
Moreover, AVSBench has two distinct sub-datasets according to the number of audio sources in a video: single-source and multi-source scenarios. 
The single-source sub-dataset is comprised of 4,932 videos, while the multi-source one contains 424 videos. 
We follow the data split of \cite{zhou2022audio} for training and testing.

\vspace{0.5 em}
\noindent
\textbf{Evaluation Metrics.} We employ the Jaccard index ($\mathcal{J}$) \cite{everingham2010pascal} and F-score ($\mathcal{F}$) as quantitative metrics for the evaluation of the predicted binary masks of acoustic objects. 
The Jaccard index, $\mathcal{J}$, is calculated as the intersection-over-union between the predicted masks and the ground-truth. 
The F-score, $\mathcal{F}$, represents the harmonic mean of precision and recall. 
This metric is expressed as $\mathcal{F}= \frac{( 1 + \beta^2)\times precision \times recall}{\beta^2 \times precision + recall}$, where $\beta^2$ is set to 0.3, as in \cite{zhou2022audio}.

\begin{table*}[t]
\centering
\caption{Quantitative comparisons with visual sounding localization methods on the Single-source and Multi-source sub-datasets. $\tau_s$ and $\tau_m$ represent the heatmap thresholds for optimal performance. 
}
\vspace{-0.5 em}
\label{table2}
\small
\setlength{\tabcolsep}{2.5mm}{%
\renewcommand{\arraystretch}{1.3} 
\begin{tabular}{ccccccccccccc}
\toprule
                          \multirow{2}{*}{Sub-dataset}  & \multicolumn{2}{c}{\begin{tabular}[c]{@{}c@{}}EZLSL~\cite{mo2022localizing}\\ ($\tau_s$=0.6, $\tau_m$=0.7)\end{tabular}} & \multicolumn{2}{c}{\begin{tabular}[c]{@{}c@{}}LVS~\cite{chen2021localizing}\\ ($\tau_s$=0.5, $\tau_m$=0.6)\end{tabular}} & \multicolumn{2}{c}{\begin{tabular}[c]{@{}c@{}}SLAVC~\cite{mo2022closer}\\ ($\tau_s$=0.5, $\tau_m$=0.6)\end{tabular}} & \multicolumn{2}{c}{\begin{tabular}[c]{@{}c@{}}SSPL~\cite{song2022self} wo/ PCM\\ ($\tau_s$=0.5, $\tau_m$=0.5)\end{tabular}} & \multicolumn{2}{c}{\begin{tabular}[c]{@{}c@{}}SSPL~\cite{song2022self} w/ PCM\\ ($\tau_s$=0.5, $\tau_m$=0.5)\end{tabular}} & \multicolumn{2}{c}{\textbf{Ours}}                                               \\ \cline{2-13} 
& $\mathcal{J}$                                      & $\mathcal{F}$                                      & $\mathcal{J}$                                     & $\mathcal{F}$                                     & $\mathcal{J}$                                      & $\mathcal{F}$                                      & $\mathcal{J}$                                          & $\mathcal{F}$                                         & $\mathcal{J}$                                         & $\mathcal{F}$                                         & $\mathcal{J}$                             & $\mathcal{F}$                             \\ \midrule
                           Single-source                 & 32.99                                  & 44.42                                  & 23.28                                 & 30.83                                 & 26.53                                  & 35.70                                  & 18.72                                      & 32.52                                     & 24.04                                     & 36.67                                     & \cellcolor[HTML]{ECF4FF}\textbf{81.29} & \cellcolor[HTML]{ECF4FF}\textbf{88.59} \\
Multi-source                  & 24.01                                  & 27.68                                  & 22.73                                 & 31.59                                 & 21.40                                  & 23.29                                  & 22.83                                      & 25.23                                     & 18.67                                     & 23.17                                     & \cellcolor[HTML]{ECF4FF}\textbf{59.5}  & \cellcolor[HTML]{ECF4FF}\textbf{65.74} \\ \bottomrule
\end{tabular}%
}
\end{table*}

\vspace{-1.0em}
\subsection{Quantitative Comparison}
\textbf{Comparison with the AVS methods.}
We compare our method with the state-of-the-art method TPAVI \cite{zhou2022audio} in Table ~\ref{table1}.
Table ~\ref{table1} indicates that our method consistently outperforms TPAVI across all the metrics. This implies our AVSC module is quite effective.
In particular, in the single-source scenario, our method achieves a 2.59\% improvement (from 78.7\% to 81.29\%) over TPAVI on the metric $\mathcal{J}$. 
In the multi-Source case, our method surpasses TPAVI by a large margin of 5.5\% on the metric $\mathcal{J}$ and achieves 59.5\% on the metric $\mathcal{F}$ which is 1.24\% higher than TPAVI. 
These results demonstrate our method performs better than the state-of-the-art on both single- and multi-source scenarios.

\vspace{0.5em}
\noindent
\textbf{Comparison with the VSL methods.}

\begin{figure}[]
	\begin{center}  
		\includegraphics[width=1.0\linewidth]{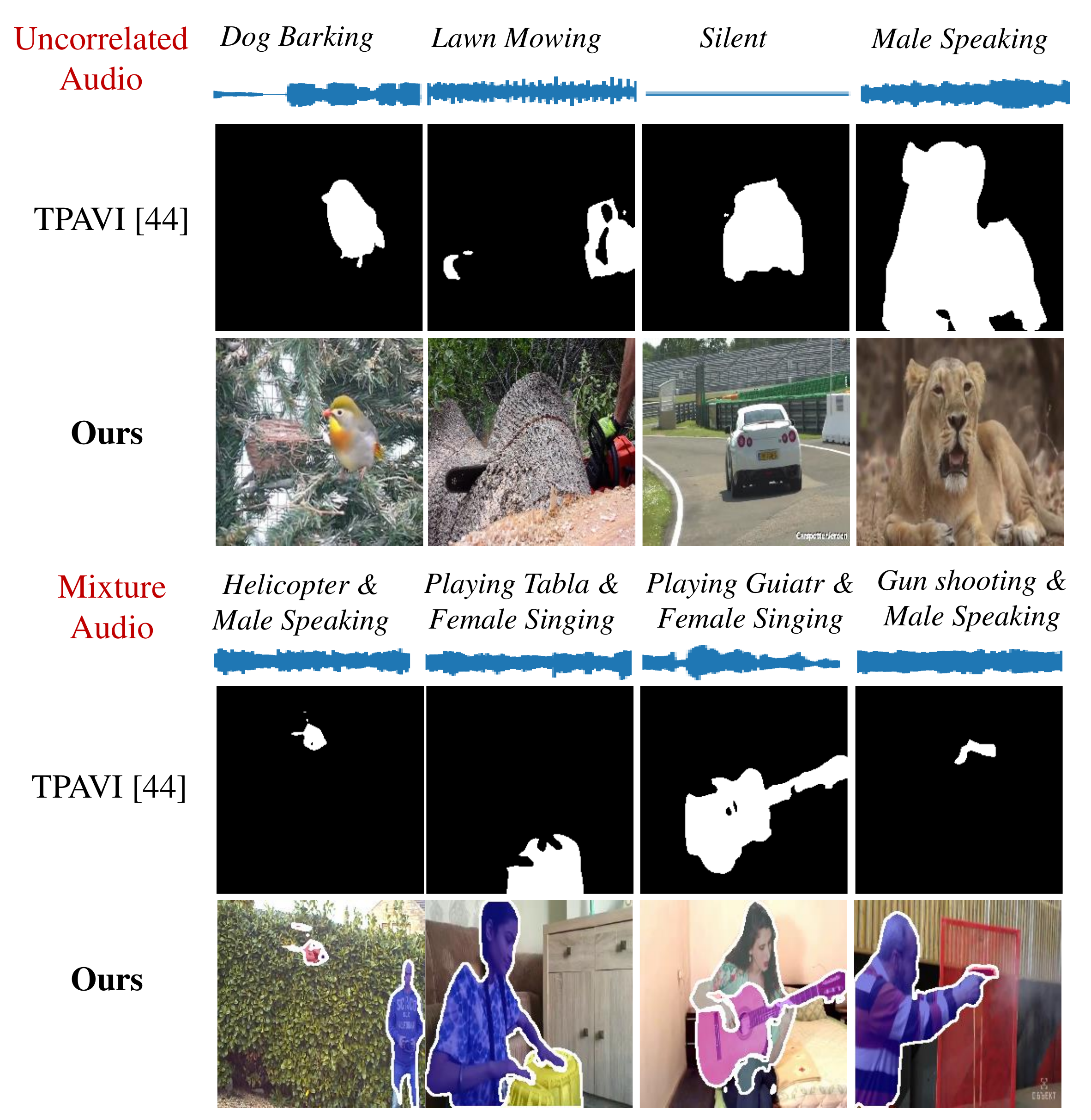} 
	\end{center}
	\vspace{-0.5em}
	\caption{Visual results of changing uncorrelated and mixed audio signals for segmentation. When there is no matching instance in a video, our method does not produce any masks.
 The original audio-visual pairs are shown in Figure ~\ref{Fig3}.}
	\label{Fig6}
 \vspace{-2.0 em}
\end{figure}

We further evaluate our approach with visual sounding localization (VSL) methods, namely EZLSL \cite{mo2022localizing}, LVS \cite{chen2021localizing}, SLAVC \cite{mo2022closer}, and SSPL \cite{song2022self}. 
For fair comparisons, we retrain EZLSL, SLAVC, and SSPL on the AVSBench dataset.
Since LVS does not release its training code, we utilize their pre-trained model for evaluation. LVS also includes all the categories of AVSBench. 
Since these methods identify sounding objects with heatmaps, we further binarize their heatmaps through a series of thresholds, ranging from 0.5 to 0.9 with intervals of 0.1. 
Then the binary masks representing the sounding objects will be used for quantitative analysis. The results are shown in Table ~\ref{table2}. 

In Table ~\ref{table2}, we present the optimal results for each approach under their respective thresholds.  
As observed, our method surpasses these VSL works by large margins.
We emphasize that our exceptional performance is primarily attributed to two designs.
First, we utilize a potential sounding object segmentation network to procure high-quality object masks.  
Second, we implement semantic association between audio and visual instances.  

\vspace{-1.0 em}
\subsection{Qualitative Comparison}

\noindent
\textbf{Comparison with the AVS method.}
Figure ~\ref{Fig3} demonstrates the results of our method and TPAVI \cite{zhou2022audio} on single audio source cases. Although both methods achieve satisfactory performance, our method performs better than TPAVI near the object boundaries.
For instance, as depicted in the penultimate column of Figure ~\ref{Fig3}, the result of TPAVI involves a portion of the hand from the gunshot sound, while our method segments the gun more precisely.

To further illustrate the effectiveness of our audio-visual semantic correlation module, we visualize the segmentation results of the multi-source cases. 
As visible in Figure ~\ref{Fig4}, even though the audio is a mixture of "Male Singing" and "Playing Piano", TPAVI only segments the most prominent object in the visual frames. 
Besides, in the multi-source setting, the segmentation masks obtained from TAPVI suffer from poor quality. 
This implies that TPAVI fails to capture the semantic correlation between the audio-visual pairs. 
In contrast, our method builds a strong semantic correlation between the audio signal and visual instances and thus achieves instance-aware sounding object segmentation.

\begin{figure}[t]
	\begin{center}  
		\includegraphics[width=0.9\linewidth]{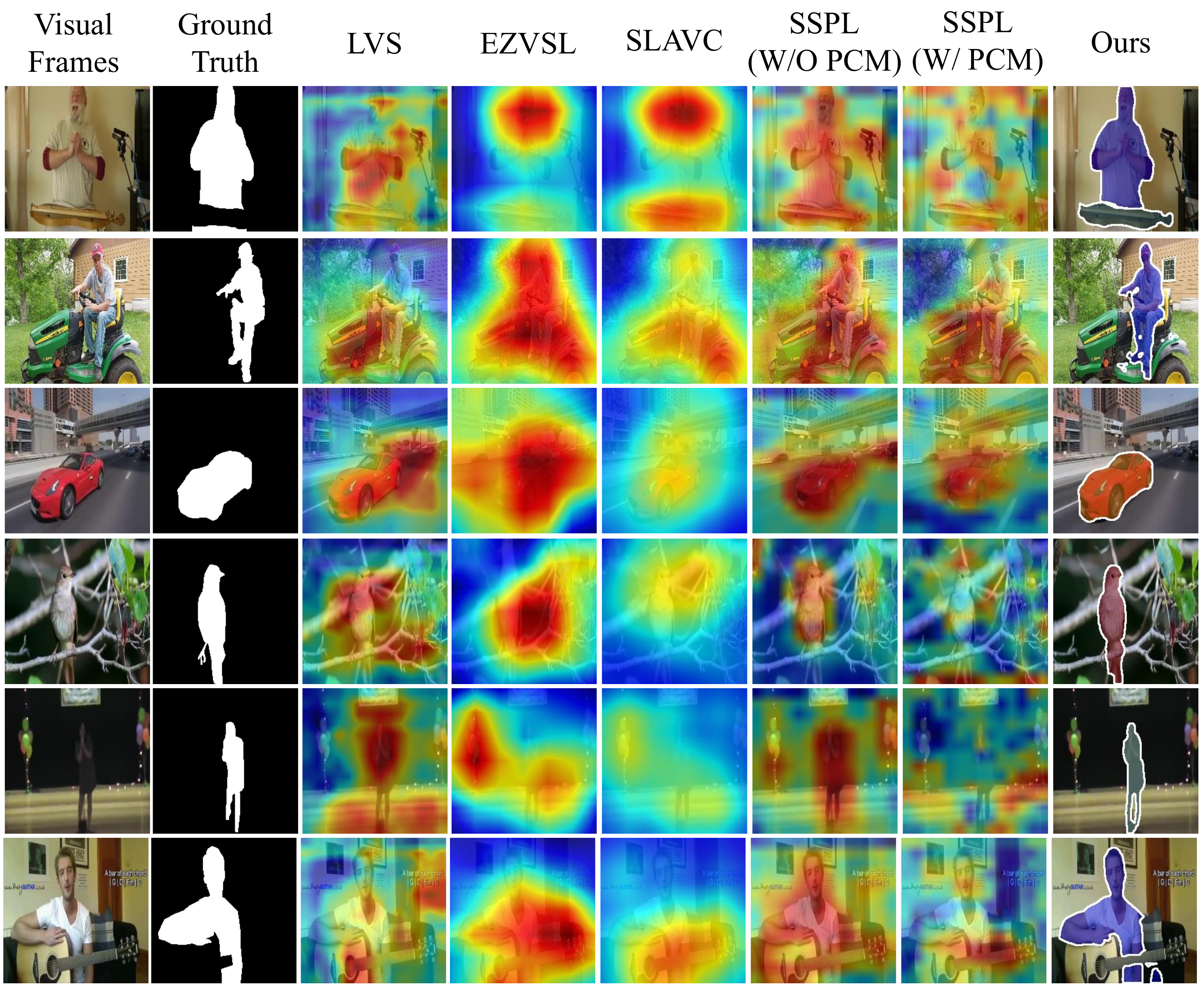} 
	\end{center}
 \vspace{-0.5em}
	\caption{Visual comparison with the visual sounding localization methods including EZLSL \cite{mo2022localizing}, LVS \cite{chen2021localizing}, SLAVC \cite{mo2022closer}, and SSPL \cite{song2022self}, respectively. }
 \vspace{-0.8em}
	\label{Fig5} 
\end{figure}

\begin{figure*}[h]
	\begin{center}  
		\includegraphics[width=0.90\linewidth]{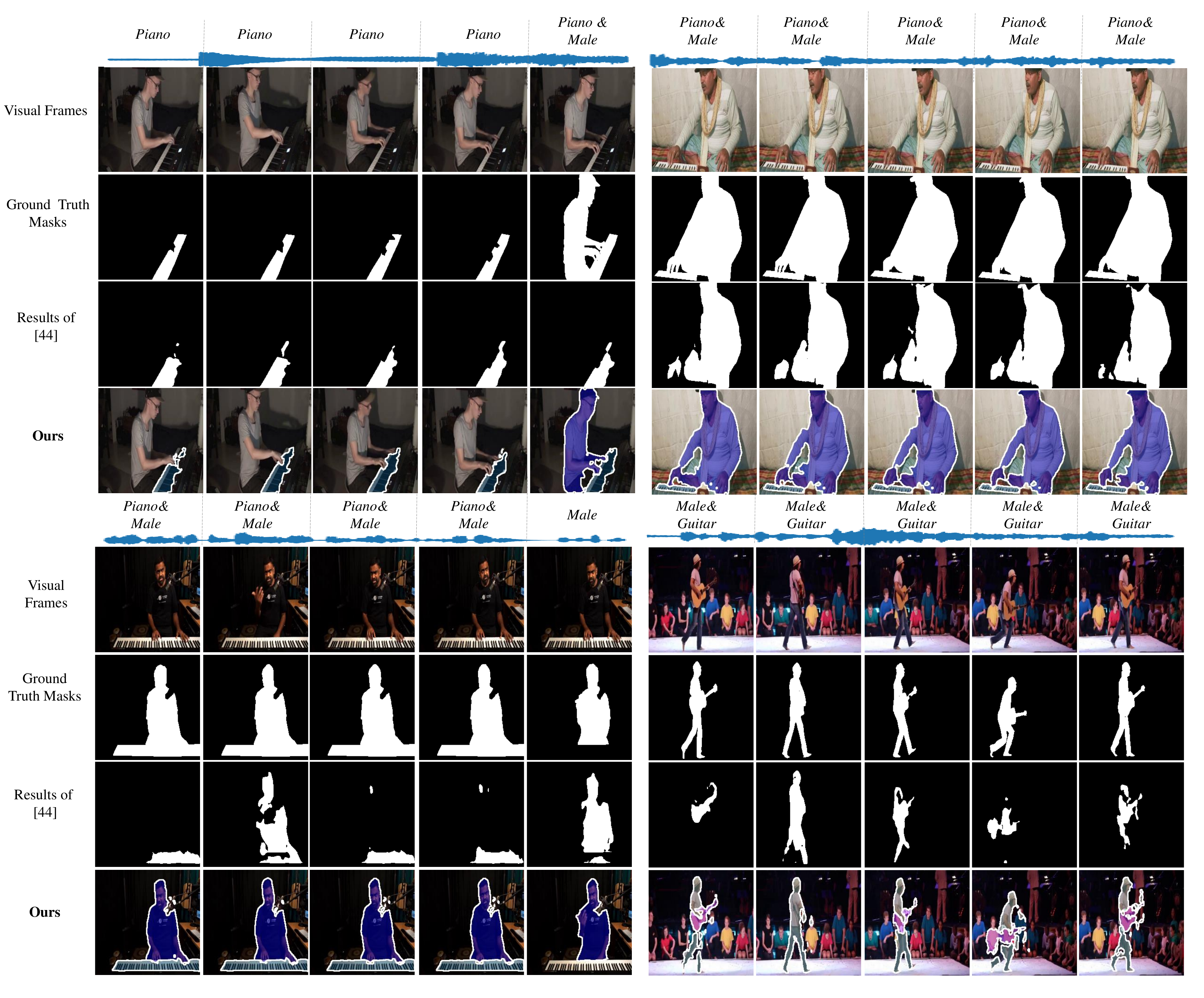}
	\end{center}
	\vspace{-2em}
	\caption{Qualitative comparisons with TPAVI \cite{zhou2022audio} on the multi-source dataset. Our method not only successfully segments sounding objects but also provides instance-level segmentation.}
	\label{Fig4} 
\end{figure*}

\noindent
\textbf{Comparison with the VSL methods.}
Figure ~\ref{Fig5} illustrates the visual results of our method and the VSL methods. 
For instance, when the input audio signal involves "Male Singing" and "Playing Guitar", VSL methods fail to localize sounding objects in a multi-source setting.
LVS and EZVSL focus on the male region while neglecting the guitar region. 
Although SSPL (\textit{w/o} PCM) emphasizes the male and piano regions, it actually puts attention on all objects within the visual frames, including the microphone. 
Similar to SSPL (\textit{w/o} PCM), SSPL (\textit{w/} PCM) also fails to localize sounding objects.
In contrast, our method effectively utilizes the audio signal to accurately segment the sounding male and guitar regions, thus achieving superior performance.

\subsection{Ablation Study}

\begin{table}[tp]
\centering
\caption{Impact of the silent object-aware segmentation loss $\mathcal{L}_{soas}$. 
$\mathcal{I}$ represents the instance number obtained from the potential sounding objects segmentation network.}
\vspace{-0.5em}
\label{tab3}
\small
\setlength{\tabcolsep}{2.3mm}{%
\renewcommand{\arraystretch}{1.3} 
\begin{tabular}{ccccccc}
\toprule
                           & \multicolumn{3}{c}{Single-source}               & \multicolumn{3}{c}{Multi-source}              \\ \cline{2-7} 
\multirow{-2}{*}{Settings} & $\mathcal{J} $ $\uparrow$               & $\mathcal{F}$ $\uparrow$             & $\mathcal{I}$ $\uparrow$              & $\mathcal{J}$ $\uparrow$             & $\mathcal{F}$ $\uparrow$             & $\mathcal{I}$ $\uparrow$           \\ \midrule
\rowcolor[HTML]{ECF4FF} 
\textbf{\textit{w/} $\mathcal{L}_{soas}$}                & \textbf{81.29} & \textbf{88.59} & \textbf{9854} & \textbf{59.5} & \textbf{65.74} & \textbf{895} \\
\textit{w/o} $\mathcal{L}_{soas}$                        & 80.67          & 88.41          & 3700          & 54.35         & 63.15          & 320          \\ \bottomrule
\end{tabular}%
}
\vspace{-1em}
\end{table}

\begin{table}[]
\centering
\caption{Impact of the AVSC on segmentation results. }
\vspace{-0.5em}
\label{tab4}
\small
\setlength{\tabcolsep}{4.2mm}{%
\renewcommand{\arraystretch}{1.3} 
\begin{tabular}{ccccc}
\toprule
                           & \multicolumn{2}{c}{Single-source} & \multicolumn{2}{c}{Multi-source} \\ \cline{2-5} 
\multirow{-2}{*}{Settings} & $\mathcal{J}$ $\uparrow$             & $\mathcal{F}$ $\uparrow$              & $\mathcal{J}$ $\uparrow$             & $\mathcal{F}$ $\uparrow$              \\ \midrule
\rowcolor[HTML]{ECF4FF} 
\textbf{\textit{w/} AVSC}   & \textbf{81.29}  & \textbf{88.59}  & \textbf{59.5}  & \textbf{65.74}  \\
\textit{w/o} AVSC          & 77.42           & 85.62           & 53.96          & 62.05           \\ \bottomrule
\end{tabular}%
}
\end{table}

\noindent
\textbf{Impact of the silent object-aware segmentation loss.} 
As discussed in Section 3.1, the majority of data in the audio-visual segmentation dataset only contains one sounding object, leading to ambiguity in training the segmentation network. 
Hence, we introduce a silent object-aware segmentation loss $\mathcal{L}_{soas}$ to enhance the variety of segmented instances. 
We perform an ablation study to investigate its impact in Table ~\ref{tab3}.
The results indicate that the silent object-aware segmentation loss improves audio-visual segmentation performance, particularly in the multi-source setting. 
Notably, the improvement of the metric $\mathcal{J}$ is nearly 5\%, and $\mathcal{F}$ increases from 63.15\% to 65.74\%. 
To further illustrate the effectiveness of $\mathcal{L}_{soas}$, we analyze the segmented instance number $\mathcal{I}$ in the test set for both single- and multi-source datasets. 
As shown in Table ~\ref{tab3}, the number of instances from the segmentation network with $\mathcal{L}_{soas}$ is three times higher than that without $\mathcal{L}_{soas}$. 
This suggests that the silent object-aware segmentation loss effectively encourages our segmentation network to segment a more diverse range of instances.

\noindent\textbf{Impact of AVSC.} To demonstrate the effectiveness of our audio-visual semantics association, we predict masks for sounding objects without using audio information. 
Then, we select the object with the highest confidence score as the final audio-visual segmentation result. 
As illustrated in Table \ref{tab4}, the performance of our method decreases significantly in the absence of audio guidance (\emph{w/o} AVSC). 
$\mathcal{J}$ decreases by 3.87\% for the single-source dataset and 5.54\% for the multi-source dataset.
This indicates that our method adequately exploits the audio signal in segmentation.

To further emphasize the sensitivity of our method to audio signals, we manually change the corresponding audio signal of the visual frames and report the visual results in Figure ~\ref{Fig6}. 
Note that, the ground truth of these audio-visual pairs can be found in Figure ~\ref{Fig3}.
As depicted in the second and third rows, when the input audio signal is unrelated to the visual frames, TPAVI still segments the object in the image, whereas our method does not segment any object. 
When we mix audio signals of two instances in a visual frame, TPAVI only segments one sounding object. 
In contrast, our method not only successfully locates two sounding objects but also achieves sound source localization at an instance level, \emph{i.e.}, associating sound sources with the corresponding objects.
 
\vspace{-1.0 em}
\section{Conclusion}

In this paper, we present an audio-visual instance-aware segmentation approach that can effectively segment sounding objects according to audio signals.
With the help of our silent object-aware segmentation training loss, we are able to segment all potential sounding objects in a video. After obtaining potential sounding objects, we can match object masks to audio signals. In other words, audio signals can modulate the segmentation results in an effective manner. Moreover, our audio-visual semantic association is instance-aware. Even though the segmentation ground-truth provided by the AVS benchmark do not provide instance-level annotation, our method can produce instance-level segmentation associated with multiple sound sources. Experiments on the popular used AVS benchmark demonstrate the superiority of our method over the state-of-the-art. More importantly, our method achieves the effect that segmentation results can be effectively adjusted by different audio signals.

\bibliographystyle{ACM_2023}
\bibliography{ACM_2023}

\newpage
\section{Appendices}

\begin{table}[h]
\caption{Comparisons with TPAVI on the inference time (Time), computation cost (GFLOPs), and parameter amount (Param).}

\label{tab:com_tpavi}
\small
\setlength{\tabcolsep}{3.5mm}{%
\renewcommand{\arraystretch}{1.5} 
\begin{tabular}{llll}
\hline
Metric         & Time (s) & GFLOPs & Param (MB) \\ \hline
TPAVI-ResNet50 & 0.0033   & 33.878 & 91.40      \\
TPAVI-PVTv2    & 0.03994  & 30.84  & 101.32     \\
Ours-ResNet50  & 0.013395 & 13.28  & 45.52      \\
Ours-PVTv2     & 0.01724  & 29.83  & 92.94      \\
Ours-Swin-B    & 0.01853  & 30.75  & 101.80     \\ \hline
\end{tabular}
}
\end{table}

\subsection{The details about matching index $\sigma$}

\noindent The matching index $\sigma$ actually represents a list that contains the index pair of the ground truth and its matching prediction.  
In other words, the list indicates which prediction is the most possible matching one with the ground truth. 
In the training stage, the matching pairs are utilized to calculate the loss, thus influencing the training effectiveness and the final results of the segmentation model.  
To be specific, incorrect matching pair causes inaccurate loss to update model parameters, degrading the model performance.  Conversely, accurate matching facilitates the learning process of our model, thus improving the model performance.

\subsection{Comparisons with AVS method}

We compare the time complexity of models.  
As suggested by Table \ref{tab:com_tpavi}, our framework shows superiority on most of the metrics, \emph{i.e.} GFLOPs and Parameter Amount.  
Note that, we also add the time for audio feature extraction into the inference time. 
Since TPAVI only provides the extracted audio features, we cannot take this time into account. 
Although our method takes longer inference time than TPAVI, it still achieves real-time interactions.  

\subsection{Ablation study about various backbones}
To demonstrate the effectiveness of our proposed module, we utilize the same backbone models as TPAVI in our framework. More specifically, we employ ResNet50 and PVT-v2 trained on ImageNet as the backbone of our segmentation model.   Concurrently, VGGish is adopted as the audio encoder for our implementation. As suggested by Table \ref{tab:tab_backbones}, our proposed method consistently achieves superior results on both metrics under the same backbone setting, indicating the effectiveness of our approach. While the improved backbone undoubtedly provides a solid foundation for the framework performance, the distinct advantage of our module should not be underemphasized.  These modules are uniquely devised to tackle the challenge of ineffective audio and to facilitate instance-level audio-visual association, as delineated in Figure \ref{Fig6}.  

\subsection{Ablation study on batchsize}
In the training stage, we utilize the model with the best performance on the validation set for the next stage of training. We re-train our framework with the consistent batch size (20) of TPAVI. As indicated by Table \ref{tab:batchsize}, under the same batch size, our method is still superior to TPAVI \cite{zhou2022audio}. For instance, on the multi-source sub-dataset, our method (ResNet50) achieves 1.7\% Jaccard Index and 3.71\% F-Score improvement over TPAVI (ResNet50). This proves that the performance improvement is mainly due to our method design rather than the batch size setting. 
\vspace{1.0em}

Table \ref{tab4} represents the ablation study for the "Audio-Visual Semantic Correlation Module" (AVSC). Note that, our proposed method aims at allowing segmentation results to be controlled by different audio signals rather than overfitting to saliency objects. To demonstrate that our method solves the problem, for the $\emph{w/o}$ AVSC, we select the segmented instances with the highest confidence score as the predicted results. Since 89\% of data in AVSBench only contain one sounding source, and the sound object is the most significant one in visual frames, Jaccard Index on both single- and multi-source datasets only increases by 3.88\% and 5.54\% separately.

\subsection{Further analysis for AVSC}
To further demonstrate the superiority of AVSC, we conduct experiments by changing the original audio signal of audio-visual pairs into the unmatching one or silent one. Under the two cases, AVS models should not segment any regions in visual frames.  Note that, different from introducing Jaccard Index and F-score to measure the quality of segmented sounding regions, we focus on whether the models still segment regions in images under the two cases, \emph{i.e.} suffer from the overfitting phenomenon.  Hence, we introduce the mIoU to measure the quality of segmented silent regions, and the recognition accuracy of silent frames or audio-visual unmatching frames (The predicted masks without any segmented sounding regions are regarded as the correct recognition.). 
\vspace{1.0em}

As suggested by Table \ref{tab:acsc}, our method achieves higher recognition accuracy of silent or unmatching audio-visual pairs while TPAVI tends to segment regions in the visual frames even for the silent cases. This further demonstrates our proposed method is sensitive to audio changes, solving the overfitting problem. 

\begin{table}
\caption{Comparisons with TPAVI under the silent audio and the unmatching audio cases. In this table, RA indicates the recognition accuracy.}

\label{tab:acsc}
\small
\small
\setlength{\tabcolsep}{2.5mm}{%
\renewcommand{\arraystretch}{1.5} 
\begin{tabular}{lccccc}
\hline
\multicolumn{2}{c}{Setting}                                              & \multicolumn{2}{c}{Single-source}               & \multicolumn{2}{c}{Multi-source}               \\ \cline{3-6} 
                                                  & \multicolumn{1}{l}{} & mIoU                    & RA  & mIoU                   & RA \\ \hline
\multicolumn{1}{c}{\multirow{2}{*}{Silent}} & TPAVI                & 81.47                   & 0.0                   & 88.34                  & 0.0                   \\
\multicolumn{1}{c}{}                              & \textbf{Ours}        & \textbf{99.00} & \textbf{94.4} & \textbf{97.79} & \textbf{80.8} \\
\multirow{2}{*}{Unmatching}                  & TPAVI                & 81.30                   & 0.0                   & 89.22                  & 0.0                   \\
                                                  & \textbf{Ours}        & \textbf{93.95} & \textbf{66.9} & \textbf{96.15} & \textbf{60.9} \\ \hline
\end{tabular}
}
\end{table}

\begin{table*}[]
\caption{Effect of the various backbones. For the segmentation network, we employ ResNet50 and PVT-v2 trained on ImageNet as the backbone.   For the audio branch, VGGish is adopted as the audio encoder for our implementation}
\label{tab:tab_backbones}
\small
\setlength{\tabcolsep}{4.5mm}{%
\renewcommand{\arraystretch}{1.2} 
\begin{tabular}{lccccc}
\hline
\multirow{2}{*}{Metric} & \multirow{2}{*}{Setting} & \multicolumn{2}{c}{TPAVI \cite{zhou2022audio}}                                               & \multicolumn{2}{c}{Ours}                                   \\ \cline{3-6} 
                        &                          & \multicolumn{1}{l}{ResNet50+VGGish} & \multicolumn{1}{l}{PVT-v2+VGGish} & ResNet50+VGGish        & \multicolumn{1}{l}{PVT-v2+VGGish} \\ \hline
\multirow{2}{*}{$J$}    & Single-Source            & 72.79                               & 78.74                             & \textbf{77.02 (+4.23)} & \textbf{80.57 (+1.83)}            \\
                        & Multi-Source             & 47.88                               & 54.00                             & \textbf{49.58 (+1.7)}  & \textbf{58.22 (+4.22)}            \\ \hline
\multirow{2}{*}{$F$}    & Single-Source            & 84.80                               & 87.90                             & \textbf{85.24 (+0.44)} & \textbf{88.19 (+0.29)}            \\
                        & Multi-Source             & 57.80                               & 64.50                             & \textbf{61.51 (+3.71)} & \textbf{65.10 (+0.6)}             \\ \hline
\end{tabular}
}
\end{table*}

\begin{table*}[]
\caption{Effects about the batchsize. In this experiment, we adopt the same batchsize as TPAVI \cite{zhou2022audio} to further demonstrate the effectiveness of our method.} 
\label{tab:batchsize}
\small
\setlength{\tabcolsep}{3.0mm}{%
\renewcommand{\arraystretch}{1.2} 
\begin{tabular}{lcccccc}
\hline
Setting                        & Metric & TPAVI (ResNet50) & Ours (ResNet50)        & TPAVI (PVT-v2) & Ours (PVT-v2)          & Ours (Swin-B) \\ \hline
\multirow{2}{*}{Single-Source} & $J$    & 72.79            & \textbf{77.02 (+4.23)} & 78.74          & \textbf{80.57 (+1.83)} & 81.12         \\
                               & $F$    & 84.80            & \textbf{85.24 (+0.44)} & 87.90          & \textbf{88.19 (+0.29)} & 88.37         \\ \hline
\multirow{2}{*}{Multi-Source}  & $J$    & 47.88            & \textbf{49.58 (+1.7)}  & 54.00          & \textbf{58.22 (+4.22)} & 59.04         \\
                               & $F$    & 57.8             & \textbf{61.51 (+3.71)} & 64.5           & \textbf{65.10 (+0.6)}  & 64.02         \\ \hline
\end{tabular}
}
\end{table*}

\end{document}